# A Quantum Good Authentication Protocol

Paul Wang

## Abstract

This article presents a novel network protocol that incorporates a quantum photonic channel for symmetric key distribution, a Dilithium signature to replace factor-based public key cryptography for enhanced authentication, security, and privacy.  The protocol uses strong hash functions to hash original messages and verify heightened data integrity at the destination.  This Quantum Good Authentication Protocol (QGP) provides high-level security provided by the theory of quantum mechanics.  QGP also has the advantage of quantum-resistant data protection that prevents current digital computer and future quantum computer attacks.

QGP transforms the Transmission Control Protocol/Internet Protocol (TCP/IP) by adding a quantum layer at the bottom of Open Systems Interconnection (OSI) model (layer 0) and modifying the top layer (layer 7) with Dilithium signatures, thus improving the security of the original OSI model.  In addition, QGP incorporates strong encryption, hardware-based quantum channels, post-quantum signatures, and secure hash algorithms over a platform of decryptors, switches, routers, and network controllers to form a testbed of the next-generation, secure quantum internet.  The experiments presented here show that QGP provides secure authentication and improved security and privacy and can be adopted as a new protocol for the next-generation quantum Internet.

## Introduction

Peter Shor sets the mathematical foundation that exponential computing problems can be reduced to polynomial on quantum computers [1–3].  Since the birth of the Shor algorithm a decade ago, many mathematicians and computer scientists have implemented and tested the algorithm in a very small scale [4–7] due to the limited qubits available on existing quantum computers.  Currently, quantum cryptanalytic algorithms can factor multiple numbers, but the mathematical forward-computing algorithms can only be effective with a huge number of error-free quantum gates.  Otherwise, the error propagation could alter the results.

Efficient quantum algorithms combined with numerical methods, error corrections, and approximation techniques can accelerate quantum Fourier transforms (QFTs), a core for the exponential speedup, so those recursive progresses could run efficiently with the limited quantum circuits [8–11].  Approximations in QFTs also improve performance in noisy intermediate scale quantum systems [12–14].

The rapidly emerging field of quantum information science and engineering has the potential to produce innovations in quantum computing, simulation, communication, sensing, and other technologies [7–11, 15–22] critical to our nation's future economic and national security.

This article introduces the design and implementation of a QGP with a Dilithium signature and quantum-enhanced communication network for improved authentication, security, and privacy to use as a testbed of the next-generation quantum internet.

## Background

### Cryptography Basics

Cryptographic algorithms have evolved over the years.  Symmetric key algorithms have the advantage of fast speed due to adopting less computationally intensive algorithms like block ciphers in the Advanced Encryption Standard (AES).  However, challenges exist in key distribution processes.

The Diffie-Hellman algorithm uses a discrete log for security.  However, it is only good for key exchanges and not for encryption or decryption, as it is vulnerable to man-in-the-middle (MiM) attacks.

Asymmetric key (public key) cryptographic systems have the advantage of easy key distribution through third-party certification authorities, but the speed is usually slow because the exponential functions used in encryption and decryption (e.g., Rivest–Shamir–Adleman [RSA]) require tremendous computational power.  In addition, RSA is integer-factoring based, which has been proven by Shor [1] to be vulnerable to attacks by future quantum computers.

Secure hash functions (e.g., SHA 1/2/3) provide the integrity for messages and are therefore commonly used for integrity verifications.

Digital signatures are used to authenticate message senders or signers, making sure a party knows the other party they are communicating with.  It is essentially the same as asymmetric key encryption, except a privacy key is applied first at signing.

There are hybrid algorithms and protocols that combine existing cryptographic techniques to provide better security and privacy.  The Pretty Good Privacy (PGP) algorithm is a well-known example.  In recent years, both quantum mechanics-based and lattice-based cryptographic and computing devices were used in cryptographic studies and applications, such as quantum random number generators (QRNGs), quantum key distribution (QKD) devices, and post-quantum cryptography (PQC) algorithms.  In the following sections, each protocol is discussed, analyzing the pros and cons of discovering a better approach (introduced in the QGP Design and Implementation section).

## RSA and Attacks on RSA

The foundation for RSA algorithms is to factor multiplication of two large prime numbers.  It is important to discover an algorithm that can reduce the computational complexity from exponential to polynomial.  **Order finding** is the core for the quantum speeding up.  The assumption is whether the execution speed can be further improved by applying numerical methods and error approximation.  Continuous analog algorithms take longer to execute for a more "precise" result, which may not be necessary since the prorogation error may exceed the gain from the deep recursive runs.

### *RSA Algorithm*

RSA is based on the following:  given a large integer *N = p × q*, it is extremely hard to find the prime factors *p* and *q*.  Let *m* (message) and *c* (ciphertext) be integers between 0 and *n* − 1 and let *e* (encryption) be an odd integer between 3 and *n* − 1 that is relatively prime to *p* − 1 and *q* − 1.  The **period** is defined as *r = (p* − 1*)(q* − 1*)*.  Without knowing *p* and *q*, it is extremely hard to find the period *r* (also known as order-finding).  The encryption/decryption applies modular exponentiation, which takes much time to compute.

### *Order Finding*

Given an integer *N* to find an integer *p* between 1 to *N* that divides *N* is not difficult.  The issue is when *N* gets very large, the exponential time needed to find a solution is beyond what classical computers can handle.  Quantum mechanics open another dimension by turning the time-consuming Fourier transforms into simple phase rotations, therefore making the nondeterministic polynomial (NP) problem solvable.  Shor's algorithm consists of two parts for the order-finding problem—a classical part that uses a reduction of the factoring problem and a quantum part that uses a quantum algorithm.  Order-finding is the key part that takes advantage of quantum computing.

The order-finding using quantum entanglement and modular exponentiation can be illustrated as follows:

1. Initialize two registers of qubits—first, an argument register with t qubits, and second, a function register with n = [*log₂N* ] bits.  These registers start in the initial state

$$|\psi_0\rangle = |0\rangle |0\rangle \quad . \tag{1}$$

2. Apply a Hadamard gate to form an equal-weighted superposition of all integers:

$$|\psi_1\rangle = \tfrac{1}{\sqrt{T}} \sum_{a=0}^{T-1} |a\rangle |0\rangle \quad . \tag{2}$$

3. Implement modular exponentiation *x<sup>a</sup> mod N* on the function registers; the state becomes

$$|\psi_2\rangle = \frac{1}{\sqrt{T}} \sum_{a=0}^{T-1} |a\rangle |x^a \mod N\rangle. \tag{3}$$

4. Perform a quantum Fourier transform on the argument register, resulting in the state

$$|\psi_3\rangle = \frac{1}{\sqrt{T}} \sum_{a=0}^{T-1} \sum_{z=0}^{T-1} e^{(2\pi)\left(\frac{aZ}{T}\right)} |Z\rangle |x^a \mod N\rangle. \tag{4}$$

## QFT And Shor Algorithm

Unlike using symbolic manipulation to solve problems as usually seen in calculus, numerical algorithms use approximation to find solutions. This is especially useful in solving engineering problems, as well as solving problems in physics, life science, medicine, financial, and aerospace. Traditional error analysis on numerical algorithms focuses mostly on discretization errors. In other words, errors caused by using a finite number of approximation of variables compared with continuous variables. Let $\omega = e^{\left(\frac{2\pi i}{T}\right)}$ and rearrange $|\psi_3\rangle$ for

$$\begin{aligned}
|\psi_3\rangle &= \frac{1}{\sqrt{T}} \sum_{a=0}^{T-1} \sum_{z=0}^{T-1} |x^a \mod N\rangle \omega^{az} |Z\rangle \\
&= \frac{1}{\sqrt{T}} \sum_{z=0}^{T-1} \sum_{a_0} |x_0^a \mod N\rangle |Z\rangle \sum_{n=0}^{T-1} \omega^{z(a_0+nr)} \\
&= \frac{1}{\sqrt{T}} \sum_{z=0}^{T-1} \sum_{a_0} |x_0^a \mod N\rangle |Z\rangle \omega^{za_0} \sum_{n=0}^{T-1} e^{\left(\frac{2\pi i}{T}\right)znr}
\end{aligned} \tag{5}$$

The phases are separated into two parts—the fixed offset $\omega^{za_0}$ and the sum of phase vectors:

$$\sum_{n=0}^{T-1} z\omega^{nr} = \sum_{n=0}^{T-1} e^{\left(\frac{2\pi i}{T}\right)znr} . \tag{6}$$

The phases spread to all directions, so most of them cancel each other out. For a few cases where $z$ is related to $r$ through $a_0$, the phases are multiples of 2π. As a result, those vectors are "amplified" as

$$\left(\frac{2\pi}{T}\right) zr = 2\pi d, \quad where \quad z = d \times \frac{T}{r} . \tag{7}$$

A QFT recursion process on quantum computers is illustrated in Figure 1. Figure 2 shows an output of a QFT, with a series of "spikes" that occur when $z$ equals to $d$ time $T$ over $r$. This can be used to calculate the period $r$.

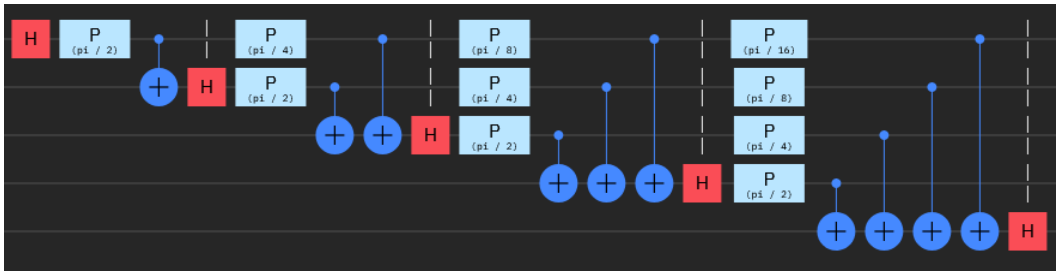

Figure 1. QFT Recursion Process. Note the θ Angle of the $e^{i\theta}$ (Phase) Component at Each Qubit Subdivides With Iterations *(Source: P. Wang)*.

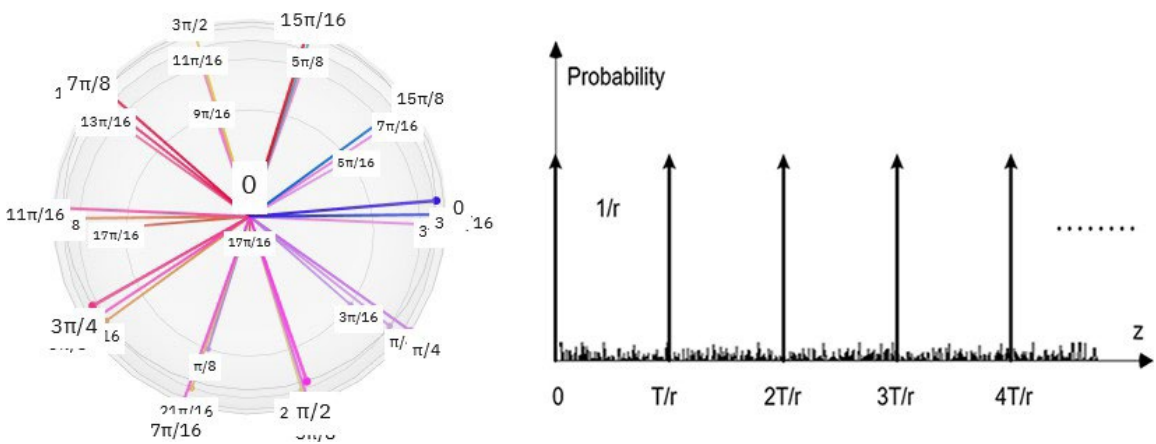

Figure 2. Quantum Phase Additions in Frequency Domain. Left: Q-Sphere Projection Looking Down Z-Axis (Top to Bottom). Right: Spikes That a Specific QFT Phase Operations Always Fall Into—Adding Up if In-Phase and Cancelling if Off-Phase *(Source: P. Wang)*.

For executing the QFT, the equation needs to run $2^{2L-1}$ times. Considering the error propagation issue, it may make sense to reduce the iterations when further recursive computations would not yield meaningful results.

## PGP

Developed by Phil Zimmerman in 1991 [23], the PGP cryptographic protocol combines symmetric, asymmetric (public key) encryptions, hashing, and digital signatures, which combine to provide the confidentiality, integrity, and secure authentication. The iconic Blackberry phone, once loved by executives, celebrities, and U.S. presidents, uses PGP end-to-end encryptions for securing emails. It was so secure that some nation states banned bringing Blackberries into their countries [24, 25]. Figure 3 shows a block diagram of the PGP algorithm, where M = message, Hm = hashed message, Epi/Dp = digital signature private key/public key, Ep/Dpi = RSA public key/private key, ZIP/Unzip = compression/decompression, and Es/Ds = symmetric key encryption/decryption.

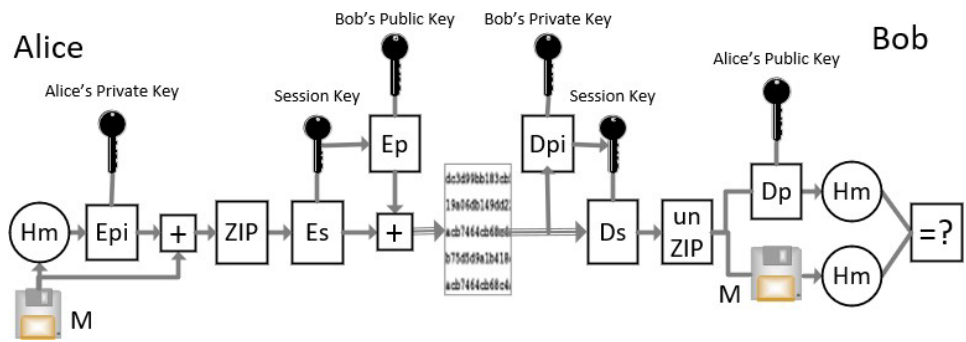

Figure 3.  PGP Algorithm Diagram  *(Source:  P. Wang)*.

Several cryptographic steps are involved in PGP. In Figure 3, the hashing of the raw message and encryption with Alice's private key is a means to provide message integrity and authentication (signature) for Bob. The raw message is appended to the hashed message, compressed (both to save transmission time and as an added layer of security), and then encrypted with a symmetric "session key." The session key is then encrypted using an asymmetric key (e.g., RSA) to share with Bob. When Bob decrypts the packet using the session key, he then recovers the hashed and raw messages. Bob then hashes the raw message himself, using the same protocol as Alice (i.e., SHA-1) so that he can compare the hash he computed with the hash value he received from Alice—if they match, then Bob knows the message is authentic (i.e., it came from Alice).

However, critical to PGP is (1) the transmission of the symmetric session key between Alice and Bob and (2) Alice's signature, which is also encrypted using public key infrastructure (RSA). The way key distribution is currently done is Alice creates a pseudorandom number (not a true random number) using the randomness of some component of her computer hardware (i.e., a mouse), encrypts the session key using Bob's public key, and sends this encrypted packet to Bob, who then decrypts it to obtain his copy of the session key.  Once Alice and Bob have copies of the session key, Bob can decrypt the raw message using asymmetric cryptography, most commonly RSA, to encrypt/decrypt the session key.  There are two issues with using asymmetric cryptography for this purpose:  (1) it is slow compared to symmetric cryptography, and (2) asymmetric cryptography (such as RSA) is a known vulnerable to quantum computers.  Note further that the hashed message is encrypted/signed using asymmetric cryptography as well as using the same public-private keys and will be vulnerable to quantum computers.  In this case, Bob would not be able to authenticate Alice.

Kuobin [26] conducted an analysis of the PGP protocol and discovered that there is a loophole in the public key distribution system that could not resist the MiM attack.  Zimmermann [27], the individual who invented the PGP, recently discussed zFone, a new encryption system for voice over IP based on PGP.  Yankubov et al. [28] proposed a blockchain-based framework for PGP key servers and showed that the distributed key management scheme can eliminate the MiM risk.  Anugurala and Chopra [29] studied secure distributed grid system infrastructure using open PGP.  Despite these efforts to make PGP more secure, the vulnerabilities to quantum computers still exist.

## PQC

Quantum computers are being developed worldwide.  Although a practical quantum computer is not yet available, basic computations on some rudimentary quantum computers have been done.  Shor's algorithm [1–3] has been proven to factor an asymmetric encryption key, such as in RSA, into its prime number components, thus exposing the public and private keys.  This significant threat to currently encrypted data is of serious concern to governments and companies, resulting in significant investment into quantum computers.  Since the digitally signed message hash also uses the factor-based public key cryptosystem, it is vulnerable to quantum attacks, most likely the "store now, decrypt later" attack.

The National Institute of Standards and Technology has recently identified four candidate PQC algorithms for standardization.  Among them, two major algorithms are recommended for most use cases—Cryptographic Suite for Algebraic Lattices (CRYSTALS)-Kyber (key establishment) and CRYSTALS-Dilithium (digital signatures) [30].  In addition, the signature schemes FALCON and SPHINCS+ are also standardized.  For this research, CRYSTALS-Kyber and CRYSTALS-Dilithium algorithms were adopted.

## CRYSTALS Kyber and Dilithium

CRYSTALS encompasses two cryptographic primitives—Kyber, a secure key-encapsulation mechanism [31], and Dilithium [32], a strongly digital signature algorithm.  Both algorithms are based on hard problems over module lattices and designed to withstand attacks by large quantum computers.  The Dilithium signature scheme  (1) is simple to implement with security, (2) is conservative with parameters, (3) minimizes the size of public keys and signatures, and

(4) is modular.  The signature scheme is composed of three steps—key generation, sign, and verification.

Although post-quantum cryptography algorithms are secure, easy to implement, and can run directly over the current TCP/IP networks, the mathematics-based, software-alone approach may face challenges in the future, as history has proved with Enigma, Data Encryption Standard (DES), and RSA.  A physics-based hardware security to complement the software approach follows the defense-in-depth security principle and adds a higher level of security to the computer networks.

## QGP Design and Implementation

To build quantum-safe networks, quantum repeaters, quantum memory, and entanglement-based key distribution systems are essential.  However, they are still under development, which is expected to take years.

Before this equipment comes into production, a secure authentication with quantum mechanics-based photonic networking devices, Dilithium-based signatures, strong encryption, and secure hash is imperative in securing today's network communications.

## Quantum Photonic Channel and Quantum Service Channel

The photonic channel uses two Cerberis XGRs linked with a dedicated fiber optic cable and a separate quantum control channel, as shown in Figure 4.

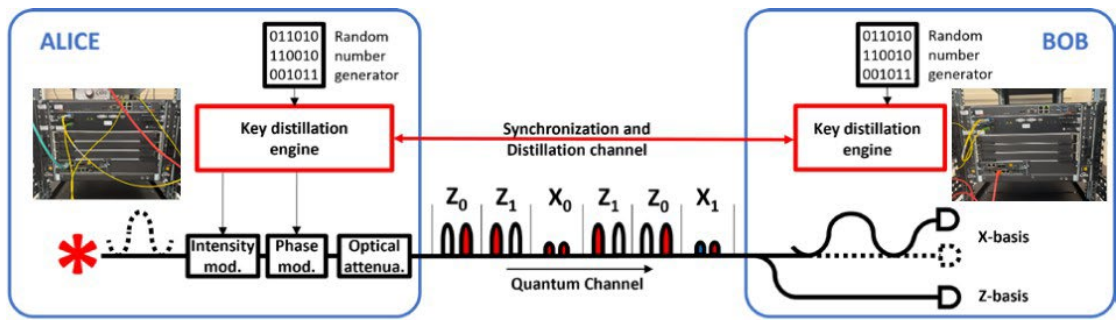

Figure 4.  Photonic Channel and Quantum Service Channel *(Source:  P. Wang)*.

The quantum service channel generates authentication keys from pure random numbers that feed into an encryptor.  Both photonic and service channels are monitored using a WebAPI Docker that runs Web services and a Simple Network Management Protocol.  The integration of the system follows the Software-Defined Network (SDN) protocol that decouples the network control logic from devices.

This sets the foundation to provide quantum keys to encryptors with a moderate key rate when needed, and devices at local and remote networks can request them.

## QGP Authentication Protocol

Security of the quantum channels of existing quantum key distribution devices depends on the assumption that both nodes are trusted.  This means both need to be hosted at locations with physical security.  In addition, security also requires "trusting" the initial handshakes.  This may be hard for notes placed at untrusted zones in countries considered adversaries.

QGP was designed and prototyped to address the security challenges in the quantum age.  It consists of a PQC-Dilithium and photonic-enhanced authentication for better security, integrity, and privacy.  A diagram of this protocol is illustrated in Figure 5, where M = message, H = hash functions (SHA 2/3), Hm = hashed message, Dis/Div = Dilithium sign/verify (PQC), ZIP/unzip = compression/decompression, Es/Ds = symmetric key encryption/decryption, and KYe/Kyd = Kyber encryption/decryption (PQC).

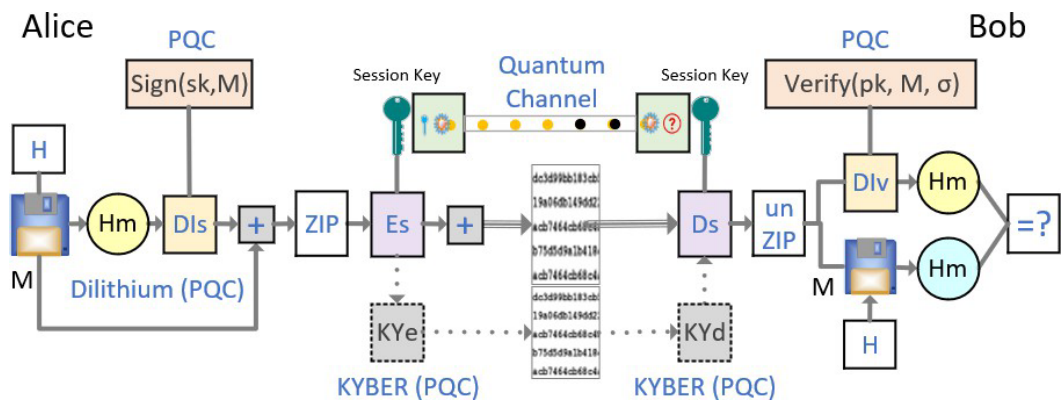

Figure 5.  QGP Authentication Protocol Diagram *(Source:  P. Wang)*.

In the QGP protocol, the plaintext (message) is hashed by Alice, and the hash value is then digitally signed (encrypted) by Dilithium (a PQC signature algorithm). The message and signed hash are then encrypted with a session/symmetric key, generated from a QKD, to be sent to Bob through a quantum photonic channel.

On the recipient side, Bob receives the session key and uses it to decrypt the message and hash and uncompress and verify the hash value.  At the same time, Bob computes a hash from the message he received.  If the received hash from Alice equals to the hash Bob computed, then message integrity is guaranteed.

In the QGP algorithm, the Dilithium signature scheme lets Bob assure the message was indeed from Alice with the PQC security.  The true random number (QRND) is used as a one-time session key.  In addition, to prevent MiM attacks, the quantum channel provides a mechanism to deliver quantum keys that cannot be observed.

KYBER (a PQC encryption/decryption algorithm) is used to encrypt and decrypt the message directly without requiring the session key. This adds another layer of security with light-weight cryptography and fast speed. A photonic chip and field-programmable gate array QGP implementation are being developed, thanks to a grant from Apple following the CHIP Act to have an advanced, onsite tape-out system for integrated circuits fabrications.

## TCP/IPQ Quantum Internet

The quantum internet is the integration of quantum computing, quantum information science, and networking.  The main advantage when comparing it with current TCP/IP networks is its use of quantum entanglement and quantum superpositions to teleport data.  Since it is based on the theory of quantum mechanics, the entangled pairs are non-observable, thus offering the highest level of security and privacy.

In addition, quantum repeaters and routers use quantum teleportation and quantum memory to store and forward data.  An ultrasecure quantum internet requires modifying the existing packet-switching network so quantum entanglement and teleportation can be integrated.

For this research, a dedicated quantum channel was added at the bottom of the physical layer of the TCP/IP protocol.  The added layer was named layer 0 and the modified protocol named TCP/IPQ.  TCP/IPQ provides not only the classical packet-switching functionality but allows entangled states to transmit in the network with less interference, thus improving security and privacy.

At the application layer (layer 7) of the TCP/IPQ protocol, the factor-based asymmetric signature was replaced with Dilithium.  The session key encryption algorithm was also replaced with Kyber.  These post-quantum cryptographic algorithms can resist quantum attacks.

Quantum-safe network stacks based on QGP and TCP/IPQ protocols are shown in Figure 6.  This is a functional prototype that can be used as a testbed for future expansions.

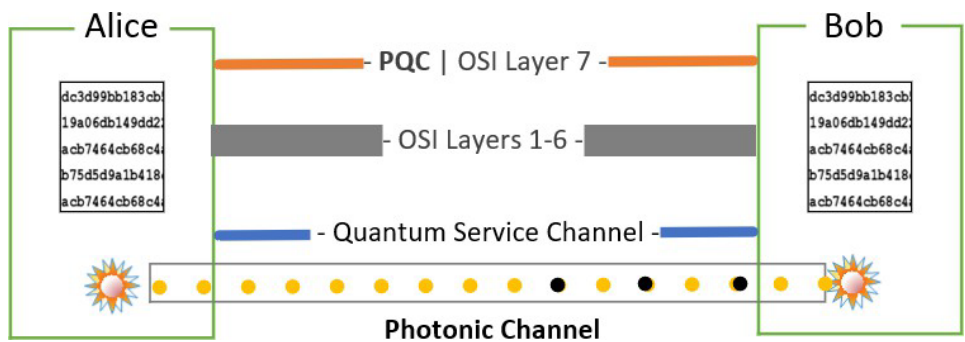

Figure 6.  TCP/IPQ Quantum Network Stacks *(Source:  P. Wang)*.

In summary, the QGP protocol consists of one photonic channel at the bottom layer of TCP/IPQ and post-quantum enabled algorithms for secure authentication.  The quantum photonic channel, two endpoints for quantum key generations, and two encryptors form the layer 0 of the TCP/IPQ.  The modified signature and symmetric key encryption at the application (top) layers of the TCP/IPQ significantly improve authentication, security, and privacy.  A quantum service channel acts as an SDN's control plane to separate the network control logic from devices.

## Experiments and Discussions

The QGP implementation integrates strong encryptions, a hardware-based quantum channel, PQC-Dilithium signatures, and secure hash algorithms to provide most secure authentication, security, and privacy.  The TCP/IPQ protocol can be used to build the next-generation quantum internet.  The experiment setting is shown in Figure 7.

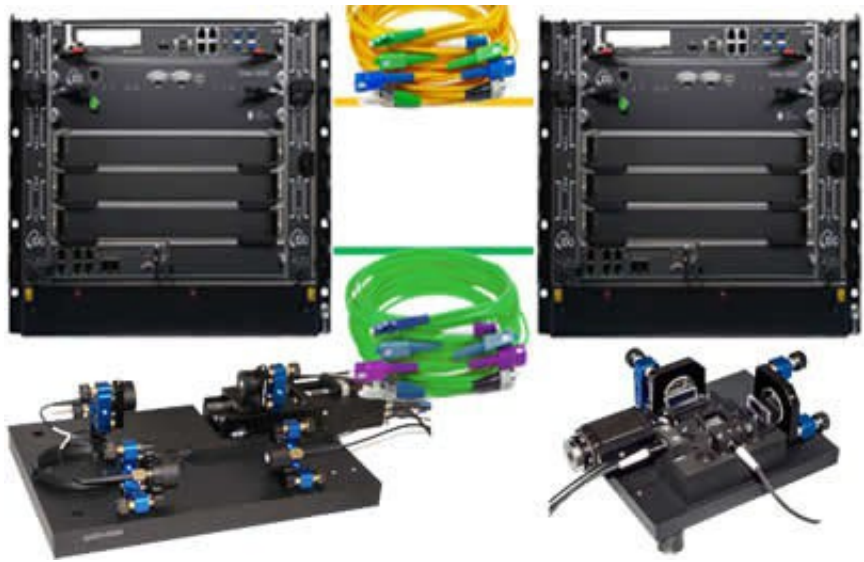

Figure 7.  TCP/IPQ Quantum Networking – Photonic Devices *(Source:  P. Wang)*.

Mao et al. conducted a study on the security aspect of the quantum photonic channel, which showed the quantum key generation, exchange, and error rates changes during a normal scenario and devices under attacks [33].  Their experiments showed error rate increased dramatically when an eavesdropping attack occurred.  More experiments, discussions, and results showing the quantum channel detecting MiM attacks can be found in Mao et al.'s article "Quantum Key Distribution and Security Studies" [33].

## Conclusions

The following is recent development in quantum networks that could benefit by the adoption of the research presented in this article.

- Although they require trusted notes, the QKD networks with more than 100 notes are being built in Europe and Asia.
- The measurement-device-independent entanglement distribution networks have shown progress by reaching hundreds of kilometers.
- The fully connected quantum network testbed can interconnect several notes.
- The quantum repeater network is emerging, and the decoy BB84 (with multiple photon intensities) has gained momentum.
- The twin-field quantum key distribution devices send photons from both ends to a central location, making it impossible for hackers to know where they came from.
- Free-space entanglements add a new dimension to quantum communication, which extends the distance between two endpoints (the endpoint can be on a satellite or the Moon).

The QGP and TCP/IPQ protocol expand the classical network protocol by adding a quantum layer (layer 0) and replacing the RSA-based algorithms with post-quantum signatures.  The novel network protocol (Figure 5) provides ultrasecure communications for today's computer networks and the future quantum internet.

## Acknowledgments

This research is funded by National Science Foundation grants #2000136 and #2329053.

## Biography

Paul Wang is a Link fellow and board member of CyberVets. He has held positions as the endowed chair in cybersecurity, director of the center for security studies, and CIO of a national organization. He was directly involved in drafting the National Initiatives of Cybersecurity Education (NICE) framework and recently published two books, with three of his patents licensed. He was advised by Dr. Robert Ledley, the inventor of the body computed topography scanner. Dr. Wang holds a Ph.D. in Computer Science and received a post-doc certificate in quantum computing from Massachusetts Institute of Technology and a certificate in artificial intelligence and data science from the University of Cambridge.